\documentclass[twocolumn,showpacs,amsmath,amssymb,superscriptaddress,aps]{revtex4-1}
\usepackage[dvips]{graphicx}
\usepackage{epsfig}
\usepackage{amsmath}
\usepackage[dvips]{color}
\newcommand{\beq}{\begin{align}}
\newcommand{\eeq}{\end{align}}

\newcommand{\cca}{\mbox{$^{12}{\rm C}+^{12}$C}}
\newcommand{\ccb}{\mbox{$^{13}{\rm C}+^{12}$C}}
\newcommand{\cnb}{\mbox{$^{13}{\rm N}+^{12}$C}}
\newcommand{\cnc}{\mbox{$^{13}{\rm N}+^{13}$C}}
\newcommand{\co}{\mbox{$^{16}{\rm O}+^{12}$C}}
\newcommand{\nc}{\mbox{$n+^{12}$C}}
\newcommand{\pc}{\mbox{$p+^{12}$C}}
\newcommand{\dbe}{\mbox{$d+^{11}$Be}}
\newcommand{\abe}{\mbox{$\alpha+^{8}$Be}}
\newcommand{\ac}{\mbox{$\alpha+^{12}$C}}

\newcommand{\ecm}{\mbox{$E_{\rm c.m.}$}}

\begin{document}
\title{Exchange effects in nucleus-nucleus reactions}
\author{J. Dohet-Eraly}
\email{jdoheter@ulb.ac.be} 
\affiliation{Physique Quantique, C.P. 165/82, Universit\'e Libre de Bruxelles (ULB), B 1050 Brussels, Belgium}
\affiliation{Physique Nucl\'eaire Th\'eorique et Physique Math\'ematique, C.P. 229, Universit\'e Libre de Bruxelles (ULB), B 1050 Brussels, Belgium}
\author{P. Descouvemont}
\email{pdesc@ulb.ac.be}
\affiliation{Physique Nucl\'eaire Th\'eorique et Physique Math\'ematique, C.P. 229, Universit\'e Libre de Bruxelles (ULB), B 1050 Brussels, Belgium}
\date{\today}
\begin{abstract}
We present a scattering model for nuclei with similar masses. In this three-body model, the projectile has
a core+valence structure, whereas the target is identical to the core nucleus. The three-body wave functions
must be symmetrized for the exchange of the cores. This property gives rise to non-local potentials, which
are computed without approximation. The present model is an extension of the Continuum Discretized Coupled Channel (CDCC)
formalism, with an additional treatment of core exchange. We solve the coupled-channel system, including non-local terms,
by the $R$-matrix method using Lagrange functions. This model is applied to the $\ccb$, $\cnb$ and $\co$ systems.
Experimental scattering cross sections are fairly well reproduced without any parameter fitting. The backward-angle
enhancement of the elastic cross sections is due to the non-local potential. We discuss in more detail the various non-local 
contributions and present effective local potentials.
\end{abstract}
\maketitle

\section{Introduction}
\label{sec1}
Nucleus-nucleus reactions represent an important topic in nuclear physics. In particular, they constitute the only way
to investigate exotic nuclei. With the development of radioactive beams, more and more data become available. Accurate
theoretical models are needed to interpret these data and to extract the relevant properties of exotic nuclei.

A popular approach is the optical model \cite{Fe58,DC19}, where the structure of the colliding nuclei is neglected.
Microscopic effects and absorption channels are simulated by complex potentials. This approach is very simple, 
but usually involves several parameters. The information about the structure of the nuclei is therefore limited.

Three-body models represent a step further in the description of nucleus-nucleus collisions. One of the participating
nuclei is described by a two-body structure, and the main part of the absorption is simulated by breakup effects in this
two-body nucleus. This approach is referred to as the Continuum Discretized Coupled Channel (CDCC) 
method \cite{Ra74,KYI86,AIK87,YMM12}, and has been extended to systems involving four-body systems \cite{MHO04,De18}. It
is well adapted to nuclei with a low separation energy, where breakup effects are expected to be important. The CDCC
method was originally developed to describe deuteron scattering \cite{Ra74}, but many applications have been
performed recently for reactions involving exotic nuclei (see, e.g., Refs.\ \cite{DAL14,PBM17} for recent works).

In its present form, the CDCC method neglects possible exchange effects between the projectile and the target. A typical
example is the $\abe$ reaction \cite{OKK09}, where the symmetrization between the colliding $\alpha$ particle and the $\alpha$'s 
involved in $^8$Be is not taken into account. A more recent example is the $\dbe$ system \cite{De17}, where $d$ and $^{11}$Be
are described by $p+n$ and $^{10}{\rm Be}+n$ structures, without antisymmetrization between the neutrons of $d$ and of
$^{11}$Be.

An obvious situation where exchange effects are important is when the colliding nuclei have similar masses. Representative
examples are the $\ccb$ and $^{17}{\rm O}+^{16}$O reactions. In such a case the system can be described by a three-body structure involving two cores
and an exchanged particle (typically a nucleon or an $\alpha$ particle). The symmetrization of the wave function for
the core exchange is then crucial. In the literature, several works have been done in this direction, with
various approximations of the exchange effects \cite{Vo70,IV87,BG66a,SBI00}.

In the present work, we use a three-body model, and treat exchange effects exactly. This procedure gives rise to
non-local potentials in a coupled-channel formalism, but does not require any parameter fit. As in the traditional
CDCC approach, the only inputs are the two-body interactions between the constituents. A first important step is to
determine the non-local potentials, stemming from exchange effects. In a second step, one has to solve a coupled-channel
integro-differential system. This is in general a complicated task, but can be simplified with the help
of the $R$-matrix formalism \cite{DB10} associated with the Lagrange-mesh technique \cite{Ba15}.

The paper is organized as follows. In Sec.\ \ref{sec2}, we present the model, with emphasis on the calculation
of the non-local terms. Section \ref{sec3} is devoted to some applications. We present results on $\ccb$, $\cnb$ and $\co$
scattering. In Sec.\ \ref{sec4}, we discuss non-local effects in more detail. We focus on the long-range
part of the non-local kernels. We also present equivalent local potentials. Concluding remarks and outlook are
presented in Sec.\ \ref{sec5}.

\section{The three-body model}
\label{sec2}

\subsection{Total wave functions}

\begin{figure}[htb]
	\begin{center}
		\epsfig{file=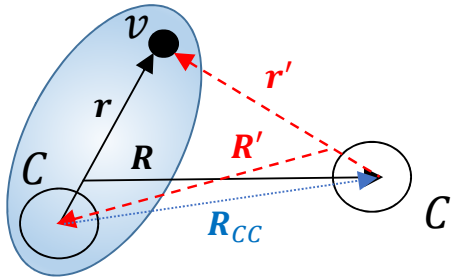,width=6cm}
		\caption{Coordinates $(\pmb{r},\pmb{R})$ and $(\pmb{r}',\pmb{R}')$. $C$ and $v$ represent the core and valence particles.}
		\label{fig_config}
	\end{center}
\end{figure}

We consider the three-body system presented in Fig.\ \ref{fig_config}.  The projectile is formed by a core ($C$) + 
valence ($v$) system, and the target is identical to the core. A typical example is 
the $\ccb$ system, where the core is $^{12}$C and the valence particle a neutron. For the sake of simplicity,
we assume that the spin of the core is zero.   The Hamiltonian of this system is defined as
\begin{align}
H=&T_{\pmb{r}}+T_{\pmb{R}}+V_{Cv}(\pmb{r})\nonumber \\
&+V_{Cv}(|\alpha \pmb{r}-\pmb{R}|)+V_{CC}(|\beta \pmb{r}+\pmb{R}|),
\label{eq1}
\end{align}
where $T_{\pmb{r}}$ and $T_{\pmb{R}}$ are the kinetic energies associated with $\pmb{r}$ and $\pmb{R}$, and $V_{Cv}$ and 
$V_{CC}$ are core-valence and core-core potentials.  The coordinates $\pmb{R}$ and $\pmb{R}'$ are the relative coordinates between the projectile and the target before and after symmetrization, respectively.
In definition (\ref{eq1}), $\alpha$ and $\beta$ are positive coefficients
given by
\begin{align}
\alpha=\frac{A_C}{A_C+A_v},\  \beta=\frac{A_v}{A_C+A_v}=1-\alpha,
\label{eq2}
\end{align}
where $A_C$ and $A_v$ are the masses of the core and of the valence particle, respectively.  We also
define
\begin{align}
\gamma=\frac{1}{1-\alpha^2},
	\label{eq2b}
\end{align}
which will be used later.

 The present approach is based on the CDCC formalism, but we include 
the symmetrization of the wave function with respect to core exchange. Notice that two possible choices exist for the
potential $V_{Cv}$: either it reproduces the spectroscopic properties of the $C+v$ system, or
it is fitted on elastic scattering. 

Let us define the core-valence Hamiltonian by
\begin{align}
H_0=T_{\pmb{r}}+V_{Cv}(\pmb{r}),
\label{eq3}
\end{align}
which is diagonalized from
\begin{align}
H_0 \phi^{\ell jm}_n(\pmb{r})=E^{\ell j}_n \phi^{\ell jm}_n(\pmb{r}).
\label{eq4}
\end{align}
The two-body wave functions are factorized as
\begin{align}
\phi^{\ell jm}_n(\pmb{r})=\frac{u^{\ell j}_n(r)}{r}
[Y_{\ell}(\Omega_r)\otimes \chi^v]^{jm},
\label{eq4b}
\end{align}
where $\chi^v$ is a spinor associated with the valence particle.
The radial eigenfunctions $u^{\ell j}_n(r)$ are expanded over a basis, chosen here from Lagrange functions \cite{Ba15}. 
Lagrange-Laguerre functions regularized by $r$ and by $\sqrt{r}$ \cite{Ba15,Do17} have been considered. Both provide the same level of accuracy.
As usual in CDCC 
calculations, energies $E^{\ell j}_n<0$ correspond to physical states, whereas states with $E^{\ell j}_n>0 $, referred 
to as pseudostates, simulate the projectile continuum.

The total wave functions, associated with (\ref{eq1}) are written before symmetrization as
\begin{align}
\Phi^{JM\pi}(\pmb{R},\pmb{r})=\frac{1}{R}\sum_{cL} g^{J\pi}_{cL}(R)\,
\varphi^{JM\pi}_{cL} (\Omega_R,\pmb{r}),
\label{eq5}
\end{align}
where index $c$ stands for $c=(\ell j n)$ and where $g^{J\pi}_{cL}$ are radial wave functions to be determined. The relative
angular momentum is denoted as $L$.
The channel functions $\varphi^{JM\pi}_{cL}$ are defined as
\begin{align}
&\varphi^{JM\pi}_{cL} (\Omega_R,\pmb{r})= 
\bigl[ 
Y_L(\Omega_R)\otimes \phi^{\ell j}_n(\pmb{r})\bigr]^{JM}.
\label{eq6}
\end{align}

The total wave function (\ref{eq5}) must be symmetrized with respect to the exchange of the cores.  For spin-zero cores,
this is achieved with the exchange operator $P$
\begin{align}
\Psi^{JM\pi}(\pmb{R},\pmb{r})=(1+P)\Phi^{JM\pi}(\pmb{R},\pmb{r}),
\label{eq7}
\end{align}
where $P$ permutes the coordinates of the cores.  More precisely, we have
\begin{align}
P\Phi^{JM\pi}(\pmb{R},\pmb{r})=\Phi^{JM\pi}(\pmb{R}',\pmb{r}'),
\label{eq8}
\end{align}
with
\begin{align}
&\pmb{r}'=\alpha \pmb{r}- \pmb{R}, \nonumber \\
&\pmb{R}'=(\alpha^2-1)\pmb{r}-\alpha \pmb{R}.
\label{eq9}
\end{align}
In the $(\pmb{R}',\pmb{r}')$ system, Hamiltonian (\ref{eq1}) can be written, in the so-called ``post" form, as
\begin{align}
&H=T_{\pmb{r}'}+T_{\pmb{R}'}+V_{Cv}(\pmb{r}')\nonumber \\
&\hspace*{1cm}+V_{Cv}(|\alpha \pmb{r}'-\pmb{R}'|)+V_{CC}(|\beta \pmb{r}'+\pmb{R}'|).
\label{eq10}
\end{align}
In the next step, we consider the three-body Schr\"{o}dinger equation
\begin{align}
H\Psi^{JM\pi}=E\Psi^{JM\pi},
\label{eq11}
\end{align}
and use (\ref{eq7}) with expansion (\ref{eq5}). After projection on the channel functions, this procedure provides the integro-differential system
\begin{align}
(T_R+E_c-E)g^{J\pi}_{cL}(R)+\sum_{c'L'}V^{J\pi}_{cL,c'L'}(R)g^{J\pi}_{c'L'}(R) \nonumber \\
+\sum_{c'L'}\int W^{J\pi}_{cL,c'L'}(R,R')g^{J\pi}_{c'L'}(R')dR'=0,
\label{eq12}
\end{align}
where
\begin{align}
T_R=-\frac{\hbar^2}{2\mu}\biggl[\frac{d^2}{dR^2}-\frac{L(L+1)}{R^2}\biggr],
\label{eq12b}
\end{align}
$\mu$ being the reduced mass.
The first two terms of Eq.~(\ref{eq12}) correspond to the standard CDCC system \cite{AIK87}. The coupling potentials are defined by
\begin{align}
V^{J\pi}_{cL,c'L'}(R)=\langle \varphi^{JM\pi}_{cL} \vert V_{Cv}+V_{CC} \vert
\varphi^{JM\pi}_{c'L'}\rangle,
\label{eq13}
\end{align}
where the integration is performed over $\Omega_R$ and $\pmb{r}$. The last term of (\ref{eq12}) is non-local, 
and arises from the symmetrization operator $P$. The non-local potential 
$W^{J\pi}_{cL,c'L'}(R,R')$ can be decomposed as
\begin{align}
&W^{J\pi}_{cL,c'L'}(R,R')=(E_c-E){\cal N}^{J\pi}_{cL,c'L'}(R,R')\nonumber \\
&\hspace*{1cm}+{\cal T}^{J\pi}_{cL,c'L'}(R,R')+
{\cal V}^{J\pi}_{cL,c'L'}(R,R'),
\label{eq14}
\end{align}
which explicitly shows overlap ${\cal N}^{J\pi}_{cL,c'L'}$, kinetic energy ${\cal T}^{J\pi}_{cL,c'L'}$ and potential 
${\cal V}^{J\pi}_{cL,c'L'}$ terms.  The overlap kernel is defined from
\begin{align}
&\int	{\cal N}^{J\pi}_{cL,c'L'}(R,R') g^{J\pi}_{c'L'}(R')dR'	= \nonumber \\
&\hspace*{1cm}	R\langle \varphi^{JM\pi}_{cL} \vert
\varphi^{JM\pi}_{c'L'}\dfrac{g^{J\pi}_{c'L'}(R')}{R'}\rangle,
\label{eqadd}
\end{align}
and equivalent expressions hold for the kinetic energy and potential kernels.
Notice that similar terms shows up in the coupled-channel approach of transfer reactions \cite{Co75}.  We discuss the various contributions of (\ref{eq14}) in the next subsection.

For scattering states ($E>0$), a radial function $g^{J\pi}_{cL}(R)$ has the asymptotic behavior at large $R$ values
\begin{align}
&g^{J\pi}_{cL,\omega L_{\omega}}(R) \rightarrow {v_c}^{-1/2}\nonumber \\
&\times \bigl[ I_L(k_c R)\delta_{c \omega} \delta_{L L_{\omega}}
-U^{J\pi}_{cL,\omega L_{\omega}}O_L(k_c R) \bigr],
\label{eq15}
\end{align}
where $\omega$ is the entrance channel, $I_L$ and $O_L$ are the incoming and outgoing Coulomb functions, $v_c$ and $k_c$
are the velocity and wave number in channel $c$,
and $\pmb{U}^{J\pi}$ is the scattering matrix.  In the $R$-matrix formalism \cite{DB10}, a channel 
radius $a$ separates the internal region, where all terms of the potentials contribute, and the external region where only 
the monopole part of the Coulomb interaction is present.  
In the internal region, the radial wave function is expanded as
\begin{align}
g^{J\pi}_{cL,\omega L_{\omega}}(R)=\sum_{i=1}^N f^{J\pi}_{cLi,\omega L_{\omega}} u_i(R),
\label{eq16}
\end{align}
where the $N$ functions $u_i(R)$ represent the basis. The choice of the basis functions is discussed in Subsec.\ \ref{sec2}.D.

\subsection{Non-local terms}
The non-local potential (\ref{eq14}) arises from exchange effects due to the operator $P$ (\ref{eq7},\ref{eq8}).  
The overlap and potential kernels in (\ref{eq14}) are obtained from
\begin{align}
&	\begin{Bmatrix}
		{\cal N}^{J\pi}_{cL,c'L'}(R,R') \\
		{\cal V}^{J\pi}_{cL,c'L'}(R,R')
		\end{Bmatrix}
={\cal J}RR'\iint  \varphi^{JM\pi\ast}_{cL}(\Omega_R,\pmb{r}) \nonumber \\
&\hspace*{1cm}\times	\begin{Bmatrix}
1 \\
V_{Cv}+V_{CC}
\end{Bmatrix}
\varphi^{JM\pi}_{c'L'}( \Omega_{R'},\pmb{r}')d\Omega_R \, d\Omega_{R'},
\label{eq18}
\end{align}
where 
\begin{align}
{\cal J}=\gamma^{3}
\end{align}
is the Jacobian from coordinates $(\pmb{r},\pmb{R})$ to $(\pmb{R},\pmb{R}')$. 
Coordinates $(\pmb{r},\pmb{r}',\pmb{R}_{CC})$ are expressed as
\begin{align}
	&\pmb{r}=-\gamma(\alpha \pmb{R}+\pmb{R}'), \nonumber \\
	&\pmb{r}'=-\gamma( \pmb{R}+\alpha \pmb{R}'), \nonumber \\
	&\pmb{R}_{CC}=\pmb{r}-\pmb{r}'=\frac{1}{\alpha+1} (\pmb{R}-\pmb{R}').
	\label{d2}
\end{align}
In the case of a nucleon transfer, $\alpha$ is close to 1, and the relative coordinates $(\pmb{r},\pmb{r}')$ are large,
even for relatively small values of $(\pmb{R},\pmb{R}')$. This means that the core-valence wave function needs to be
accurately known up to large distances. If the binding energy is small, the non-local potentials are therefore
very sensitive to the long-range part of bound-state wave function.

The potential term is quite similar to the matrix elements involved in Distorted Wave Born Approximation
(DWBA) calculations \cite{Sa83}, and its calculation 
is explained in several references (see e.g. Refs.\ \cite{Sa83,Th88,SD19}).  The calculation of the overlap
and potential kernels is based on the expansions
\begin{align}
&\frac{u^{\ell j}_n(r)\, u^{\ell' j'}_{n'}(r')}{r^{\ell+1}r'^{\ell'+1}}=
\sum_K N^{K}_{cc'}(R,R')P_K(\cos \theta_R) ,\nonumber \\
&\frac{u^{\ell j}_n(r)}{r^{\ell+1}}
\biggl(V_{Cv}(r')+V_{CC}(R_{CC})\biggr) \frac{u^{\ell' j'}_{n'}(r')}{r'^{\ell'+1}}= \nonumber \\
&\hspace*{1cm} \sum_K V^{K}_{cc'}(R,R')P_K(\cos \theta_R) ,
\label{eq19}
\end{align}
where $\theta_R$ is the angle between $\pmb{R}$ and $\pmb{R}'$, and $P_K(x)$ is a Legendre polynomial. The components 
$N^{K}_{cc'}(R,R')$ and $V^{K}_{cc'}(R,R')$ are obtained from numerical integrations.  Notice that these expansions only depend on the quantum numbers of the projectile. They do not depend on the angular momenta $J,L,L'$, and 
are therefore performed once.
 
The derivation of the kinetic-energy kernels is more tedious.  However, we show in the Appendix, that they can be 
deduced from the overlap kernel as
\begin{align}
{\cal T}^{J\pi}_{cL,c'L'}(R,R')=-\frac{\hbar^2}{2\mu}
\biggl[\frac{\partial^2}{\partial R^2}-\frac{L(L+1)}{R^2}\biggr]
{\cal N}^{J\pi}_{cL,c'L'}(R,R').
\label{eq20}
\end{align}

The calculation of the non-local kernels from (\ref{eq18}) and (\ref{eq19}) requires some angular-momentum algebra. 
We use the addition theorem
\begin{align}
	r^{\ell} Y_{\ell}^m (\Omega_r)=\sum_{\lambda=0}^{\ell} C_{\ell}^{\lambda}r_1^{\lambda}r_2^{{\ell}-{\lambda}}
	\bigl[Y_{\lambda}(\Omega_1)\otimes Y_{{\ell}-{\lambda}}(\Omega_2)\bigr]^{\ell m},
	\label{eq20b}
\end{align}
where $\pmb{r}=\pmb{r}_1+\pmb{r}_2$ and with
\begin{align}
C_{\ell}^{\lambda}=\biggl[\frac{4\pi (2\ell+1)!}{(2\lambda+1)!(2\ell-2\lambda+1)!}\biggr]^{1/2}.
\end{align}
When the valence particle $v$ has a spin zero, the overlap kernel can be expanded as
\begin{align}
&{\cal N}^{J\pi}_{cL,c'L'}(R,R')={\cal J}(-\gamma)^{\ell+\ell'}\sum_K
N^K_{c,c'}(R,R') \nonumber \\
&\times \sum_{\lambda_1 \lambda_2} C_{\ell}^{\lambda_1}C_{\ell'}^{\lambda_2}\alpha^{\lambda_1+\lambda_2}
R^{\ell'+\lambda_1-\lambda_2+1}R'^{\ell-\lambda_1+\lambda_2+1} \nonumber \\
&\times F^{J\pi}_{cL,c'L'}(K,\lambda_1,\lambda_2),
\end{align}
with 
\begin{align}
& F^{J\pi}_{cL,c'L'}(K,\lambda_1,\lambda_2)=\nonumber \\
&\biggl\langle \biggl[ Y_{L}(\Omega_R)\otimes
\bigl[Y_{\lambda_1}(\Omega_R)\otimes Y_{\ell-\lambda_1}(\Omega_{R'})\bigr]^{\ell}\biggr]^{JM} \vert
P_K(\cos \theta_R)\vert \nonumber \\
& \biggl[ Y_{L'}(\Omega_{R'})\otimes
\bigl[Y_{\ell'-\lambda_2}(\Omega_R)\otimes Y_{\lambda_2}(\Omega_{R'})\bigr]^{\ell'}\biggr]^{JM}
\biggr\rangle.
\end{align}
The analytical calculation of these coefficients requires some algebra to modify the order of angular-momentum
couplings, and involves $6j$ coefficients. When the valence particle has a spin, further angular-momentum recoupling
is necessary. A simple value is obtained for $\ell=\ell'=0$, where we have
\begin{align}
F^{J\pi}(K,0,0)=\frac{\delta_{KJ}}{4\pi(2J+1)}.
\end{align}

\subsection{Symmetry of the non-local kernels}
Let us briefly discuss the symmetry properties of the non-local potential (\ref{eq14}).  According to Eq.\ (\ref{eq7}), 
we have, when the cores are bosons
\begin{align}
P\Psi^{JM\pi}=\Psi^{JM\pi},
\label{eq21}
\end{align}
which means that the property
\begin{align}
[H,P]=0
\label{eq22}
\end{align}
should be satisfied.  This implies that both $V_{Cv}$ potentials in Eqs.\ (\ref{eq1}) and (\ref{eq10}) are identical.  For example, 
in the $\ccb$ system, the $\nc$ potential (real) associated with the $^{13}$C ground state should be identical to 
the $\nc$ optical potential which describes the neutron-target scattering.  If this condition is fulfilled,  
the non-local potential $W^{J\pi}$ is symmetric, and we have
\begin{align}
W^{J\pi}_{cL,c'L'}(R,R')=W^{J\pi}_{c'L',cL}(R',R).
\label{eq23}
\end{align}
This test is very strong since, individually, all terms of the r.h.s.\ of (\ref{eq14}) are not symmetric.  
In practical applications, however, it may seem more physical to choose different potentials: one which binds 
the $C+v$ system, and the other which is adapted to the $C+v$ scattering.  In such a case, the symmetry 
property (\ref{eq23}) is approximately satisfied (see the discussion in Ref.\ \cite{IV87}).  This means that some properties 
of the scattering matrix, such as the symmetry or the unitarity (for real potentials), are not any more valid.  
We choose to restore the symmetry of the non-local potential through
\begin{align}
W^{J\pi}_{cL,c'L'}(R,R')\rightarrow \frac{1}{2}\biggl[
W^{J\pi}_{cL,c'L'}(R,R')+W^{J\pi}_{c'L',cL}(R',R) \biggr].
\label{eq24}
\end{align}

We will see in some examples that these effects, in practice, are small.  The main reason is that, in general, the contribution of
the optical potential $V_{Cv}$ is small compared to the optical potential between the cores.

\subsection{Elastic cross sections}
The elastic cross sections are obtained from the scattering matrices $\pmb{U}^{J\pi}$ [see Eq.\ (\ref{eq15})].
According to the scattering theory, the elastic cross section between different particles is defined
from the scattering amplitude as
\begin{align}
	&\frac{d\sigma}{d\theta}=\vert f(\theta)\vert ^2,\nonumber \\
	&f(\theta)=f^{\rm N}(\theta)+f^{\rm C}(\theta),
\label{eqc1}
\end{align}
where $\theta$ is the scattering angle.
In this definition, $f^{\rm C}(\theta)$ is the Coulomb amplitude and $f^{\rm N}(\theta)$ is the nuclear amplitude, defined by
\begin{align}
f^{\rm N}(\theta)=\frac{1}{2ik}\sum_J (2J+1)P_J(\cos \theta) (U^J-1)e^{2i\sigma_J},
\label{eqc2}
\end{align}
where $\sigma_J$ is the Coulomb phase shift. For the sake of simplicity, we consider single-channel 
systems with spin 0 nuclei. The 
generalization is straightforward (see, for example, Ref.\ \cite{Sa83}).

As explained in Sec.\ \ref{sec2}.A, the scattering matrices are obtained from the resolution of a Schr\"{o}dinger
equation involving a non-local potential. Let us denote as $U^J_0$ and $g^J_0(R)$ the scattering matrix and
wave function obtained from the local term (\ref{eq13}) only. These quantities are obtained without symmetrization
of the wave function (\ref{eq7}). The integral definition of the scattering matrix \cite{CH13} provides
a relationship between $U^J$ and $U^J_0$ as
\begin{align}
&U^J =U^J_0+U^J_{\rm ex},\nonumber \\
&U^J_{\rm ex}=-\frac{i}{\hbar}\iint g^J_0(R) W^J(R,R')g^J(R') dR dR'.
\label{eqc3}
\end{align}
Consequently the nuclear scattering amplitude can be written as
\begin{align}
f^{\rm N}(\theta)=f^{\rm N}_0(\theta)+f^{\rm N}_{\rm ex}(\theta),	
\label{eqc4}
\end{align}
where $f^{\rm N}_0(\theta)$ is obtained from (\ref{eqc2}) with the scattering matrices $U^J_0$, and where the exchange amplitude
$f^{\rm N}_{\rm ex}(\theta)$ is the non-local contribution
\begin{align}
f^{\rm N}_{\rm ex}(\theta)=	\frac{1}{2ik}\sum_J (2J+1)P_J(\cos \theta) U^J_{\rm ex}e^{2i\sigma_J}.
\label{eqc5}
\end{align}
A similar decomposition has been suggested in Refs.\ \cite{PPK18,PKK21,PMP19}. 
Since the calculation of the exchange amplitude $f^{\rm N}_{\rm ex}(\theta)$ is based on a non-local
potential, it is common in the literature to
assume that it can be simulated by the transfer of the valence particle (see for example Ref.\ \cite{PMP19} 
and references therein). In other words, the exchange amplitude
can be approximated as
\begin{align}
	f^{\rm N}_{\rm ex}(\theta)\approx f_{\rm tr}(\pi-\theta),
	\label{eqc6}
\end{align}
where $f_{\rm tr}(\pi-\theta)$ is the transfer amplitude, usually computed at the DWBA approximation \cite{Th88}.
In this approximation, however, the exact wave function $g^J(R)$ is replaced by the non-symmetrized
wave function $g^J_0(R)$ in Eq.\ (\ref{eqc3}), and the non-local potential $W^J(R,R')$ is approximated
from an auxiliary potential.

Another difference is related to the spectroscopic factors. As expected in transfer calculations, the DWBA
amplitude $f_{\rm tr}(\pi-\theta)$ is multiplied by the spectroscopic factor of the projectile. This
multiplicative factor, however, does not show up in the present approach. The two-body wave function
(\ref{eq4b}) is of course an approximation which can be improved, either by introducing a spectroscopic
factor or by including core excitations. The introduction of a spectroscopic factor in (\ref{eq4b}), however,
represents a global renormalization of the expansion (\ref{eq5}), and therefore of the symmetrized
definition (\ref{eq7}). The scattering matrices deduced from (\ref{eq15}) are therefore not affected by
a spectroscopic factor. Introducing core excitations in the present approach is a challenge for
future works, but is beyond the scope of the present work.

\subsection{Lagrange functions}
As mentioned before, the scattering matrices are calculated with the $R$-matrix method, which is based on a
channel radius $a$ and on the choice of basis functions $u_i$. As in previous works, we choose Lagrange functions
which permit fast and accurate calculations of the matrix elements, in particular for non-local potentials (see Ref.\ \cite{Ba15} for detail).  

The calculation of the $R$ matrix is based on matrix elements between basis functions over the internal region. The main input is the matrix defined from
\begin{align}
C^{J\pi}_{cLi,c'L'j}=&\langle u_i \vert
(T_R+E_c-E)\delta_{cc'}\delta_{LL'}\nonumber \\
&+V^{J\pi}_{cL,c'L'}+ W^{J\pi}_{cL,c'L'}
\vert u_j \rangle.
\label{eql1}
\end{align}
For example, matrix elements of the local and non-local potentials are given by
\begin{align}
&\langle u_i \vert V \vert u_j \rangle=\int_0^a u_i(R)V(R)u_j(R)\, dR ,\nonumber \\
&\langle u_i \vert W \vert u_j \rangle=\int_0^a\int_0^a u_i(R)W(R,R')u_j(R')\, dR dR'.
\label{eql2}
\end{align}
These calculations are greatly simplified by using Lagrange functions for $u_i$ which are defined by
\begin{align}
u_i (R)=(-1)^{N+i} \frac{R}{R_i}   \sqrt{R_i\biggl(1-\frac{R_i}{a}\biggr)}\,
\frac{P_N(2R/a-1) }{R-R_i},
\label{eql3}
\end{align}
where $R_i$ are the zeros of
\begin{align}
P_N(2R/a-1)=0.
\label{eql4}
\end{align}
The normalization of (\ref{eql3}) is chosen in such a way that the Lagrange condition
\begin{eqnarray}
u_i(R_j)=\frac{1}{\sqrt{a\lambda_i}}\delta_{ij}
\label{eql5}
\end{eqnarray}
is satisfied.  In this equation, $\lambda_i$ is the weight of the Gauss-Legendre 
quadrature associated with the $[0,1]$ interval.

With the choice of basis function (\ref{eql3}), the calculation of the matrix elements (\ref{eql2})
is extremely simple if the Gauss approximation of order $N$ is used for the quadratures. The matrix elements are given by
\begin{align}
&\langle u_i \vert V \vert u_j \rangle=V(R_i)\delta_{ij}, \nonumber \\
&\langle u_i \vert W \vert u_j \rangle=a\sqrt{\lambda_i \lambda_j}W(R_i,R_j),
\end{align}
and no numerical integral is required for the matrix elements. We refer to Refs.\ \cite{DB10,Ba15} for details.

\section{Applications}
\label{sec3}
\subsection{The $\ccb$ system}
The $\ccb$ system has been intensively studied experimentally \cite{LBD95,VBT88} as well as theoretically \cite{Vo70,IV87,IDM97}.
It is known that non-local effects can be simulated by a parity-dependent optical potential \cite{BG66a}. Owing to the
one-neutron exchange, the potentials for even and odd partial waves are different \cite{Ba86}. In the ``extreme" situation of identical
nuclei, odd partial waves are strictly forbidden.

We have determined the non-local potential (\ref{eq14}) from $\cca$ and $\nc$ potentials. For $\cca$, we take the optical
potential derived by Treu {\sl et al.}\ \cite{TFG80}, and defined (in MeV) as
\begin{align}
V_{\rm CC}(r)=&-\frac{100}{1+\exp[(r-5.45)/0.48]}\nonumber \\
&-i\frac{15}{1+\exp[(r-5.77)/0.26]}
\label{eq25}
\end{align}
where $r$ is expressed in fm. A Coulomb point-sphere potential of radius $R_C=5.45$ fm is added. This optical
potential is fitted on elastic-scattering data around the Coulomb barrier. In all applications,
we use the integer masses with $\hbar^2/2m_N=20.736$ MeV$\,$fm$^2$ ($m_N$ is the nucleon mass). 

The $\nc$ potential is chosen as in Ref.\ \cite{VI84}, i.e.
\begin{align}
V_{nC}(\pmb{r})=&-\frac{V_0}{1+\exp[(r-r_0)/a_0]}\nonumber \\
&-(\pmb{\ell}\cdot\pmb{s})\frac{V_{\ell s}}{r}\frac{d}{dr}\frac{1}{1+\exp[(r-r_0)/a_0]},
\label{eq26}
\end{align}
where $V_0=62.70$ MeV for $\ell$ even and $50.59$ MeV for $\ell$ odd, and where $r_0=2.656$ fm and $a_0=0.705$ fm.  
The spin-orbit amplitude is $V_{\ell s}=28.406$ MeV.  This potential reproduces the experimental energies of the 
first $1/2^-,1/2^+$, and $5/2^+$ states in $^{13}$C.  Between the target and the projectile, the same $\nc$ potential is adopted for each partial wave, corresponding to the central part of (\ref{eq26}). 
Once potentials (\ref{eq25}) and (\ref{eq26}) are determined, 
the model does not contain any free parameter.

\begin{figure}[htb]
	\begin{center}
		\epsfig{file=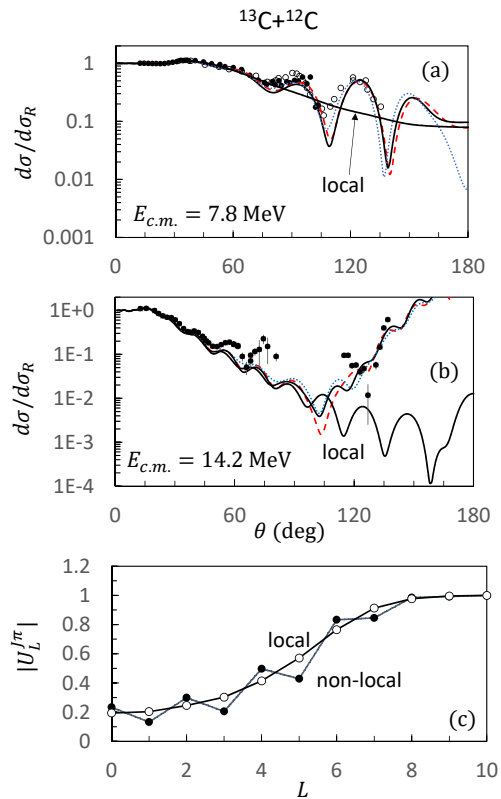,width=6.5cm}
		\caption{$\ccb$ elastic cross sections (divided by the Rutherford cross section) at $\ecm=7.8$ MeV (a) and 14.2 MeV (b).
			The data are taken from Refs.\ \cite{LBD95} (full dots) and \cite{VBT88} (open dots). The solid lines are
			obtained with the $1/2^-,1/2^+,5/2^+$ states of $^{13}$C. The dashed (red) lines correspond to the $1/2^-$ ground state only. The dotted (blue) lines are obtained with different $\nc$ potentials in the entrance and exit channels (see text).
			Panel (c) presents the amplitudes $\vert U^{J\pi}_L\vert$ of the scattering matrices at $\ecm=7.8$ MeV, and for $J=L-1/2$.}
		\label{fig_c12c13_ela}
	\end{center}
\end{figure}

In Fig.\ \ref{fig_c12c13_ela}, the $\ccb$ elastic cross sections at $\ecm=7.8$ and 14.2 MeV are shown.  In each case, we 
consider four conditions: (1) when only the local potential is included, the backward angle enhancement of the 
cross section is not reproduced; (2) the non-local calculation involving the $^{13}$C ground state only (dashed line) reproduces fairly 
well the data; (3) when the $1/2^+$ and $5/2^+$ excited states are introduced (solid line), elastic scattering is not 
significantly modified; (4) in the $\nc$ potential between the target and the projectile (dotted line), we have replaced the real 
potential (\ref{eq26}) by the Koning-Delaroche parametrization \cite{KD03}.  Although some symmetry properties 
are lost (see Sec.\ \ref{sec2}.B), there is a weak influence on the $\ccb$ cross section.

Figure \ref{fig_c12c13_ela}(c) presents the amplitudes $\vert U^{J\pi}_L\vert$ of the scattering matrices 
(elastic channel) at  $\ecm=7.8$ MeV.  We choose here $J=L-1/2$ but a similar behaviour is observed for $J=L+1/2$.  Without 
the non-local part of the potential, the variation is smooth.  As expected, the non-locality leads to a splitting 
between odd and even $L$ values.  This property gives rise to the backward angle enhancement of the cross section, 
and justifies the use of parity-dependent optical potentials \cite{BG66a} to simulate non-local effects.  
An advantage of the present method is that 
it does not require any additional parameter.  In addition, excited states of the $\nc$ system are included 
in a straightforward way.

\begin{figure}[htb]
	\begin{center}
		\epsfig{file=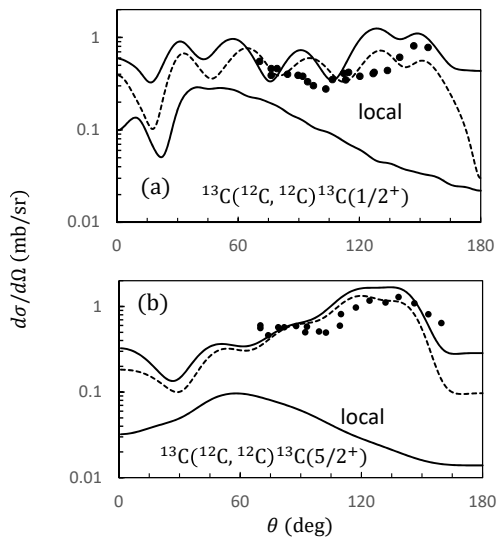,width=6.5cm}
		\caption{Inelastic $\ccb$ cross sections to the $1/2^+$ (a) and $5/2^+$ (b) states of  $^{13}$C at $\ecm=9.88$
			MeV. The experimental data are taken from Ref.\ \cite{FBT84}. Solid (dashed) lines are obtained with
		identical (different) $\nc$ potentials in the entrance and exit channels (see text).}
		\label{fig_c12c13_inela}
	\end{center}
\end{figure}

In Fig.\ \ref{fig_c12c13_inela}, we investigate the inelastic cross sections to the $^{13}$C($1/2^+$) and
$^{13}$C($5/2^+$) states, which have been measured in Ref.\ \cite{FBT84} at energies around the Coulomb barrier. The effect of
non-locality is quite important. With the local potential only, the theoretical cross sections are far below
the data. The calculation involves the $^{13}{\rm C(gs},1/2^+,5/2^+)+^{12}$C channels. Here the cross sections are more sensitive
to the choice of the $\nc$ potential. Let us emphasize that there is no fit of the cross sections. All inputs are
kept identical as for elastic scattering.

\subsection{The $\cnb$ system}
\begin{figure}[htb]
	\begin{center}
		\epsfig{file=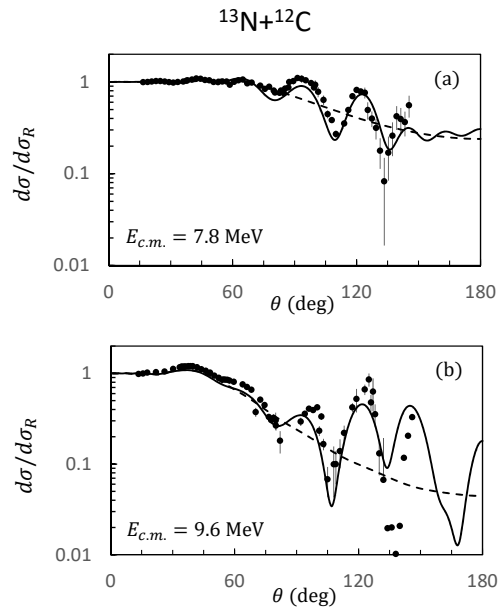,width=6.5cm}
		\caption{$\cnb$ elastic cross sections (divided by the Rutherford cross section) at $\ecm=7.8$ (a) and 9.6 (b) MeV.
			The data are taken from Ref.\ \cite{LBD95}. Solid lines correspond to the full calculation, and dashed lines to the
		local potential only.}
		\label{fig_c12n13_ela}
	\end{center}
\end{figure}

With the development of radioactive beams, the $\cnb$ mirror system has attracted much attention 
in the literature \cite{LBD95,IDM97}. Measurements have
provided some information about charge-symmetry
and about the parity effect \cite{LBD95}. Figure \ref{fig_c12n13_ela} shows the $\cnb$ calculated cross sections. 
Only the $^{13}$N ground state is bound, and has been introduced in the calculation. With respect to the $\ccb$ system, 
the only difference is the introduction of a Coulomb term for $\pc$ (with $R_C=2.7$ fm).  
The binding energy of $^{13}$N
is $-1.90$ MeV, in fair agreement with experiment ($-1.94$ MeV). 
Here again the model reproduces remarkably well the experimental data \cite{LBD95}.

\subsection{The $\co$ system}
In the $\co$ scattering, an $\alpha$ particle is exchanged between the target and the projectile. This system has
been intensively investigated in the literature (see, for example, Refs.\ \cite{PMP19,PPK18,PKK21} for recent works). 

We consider the elastic-scattering data of Villari {\sl et al.} \cite{VLL89} at the typical energy $\ecm=23.14$ MeV, where
a backward-angle enhancement of the cross sections is observed. We use the same $\cca$ core-core potential (\ref{eq25})
as in previous applications. For the
$\ac$ system, we adopt the potentials used in Ref.\ \cite{OHR12}, i.e.\ a Woods-Saxon potential with a range $R_0=4.15$ fm
and a diffuseness $a=0.55$ fm. In addition to the $0^+$ ground state, we include the $1^-,3^-$, and $2^+$ excited states.
Since the $2^+$ state presents a cluster structure, $R_0$ is chosen larger (4.5 fm) for this state. 
The Coulomb potential has a point-sphere shape
with a radius $R_C=4.15$ fm. The depths of the potentials $V_0$ are adjusted to the experimental
binding energies, which provides $V_0=-43.25,-68.95, -41.3$, and $-42.4$ MeV for the $0^+,2^+,1^-$, and $3^-$ states, respectively. Between the target and the projectile, the same $\ac$ potential is adopted, corresponding
to the $^{16}$O ground state.

\begin{figure}[htb]
\begin{center}
\epsfig{file=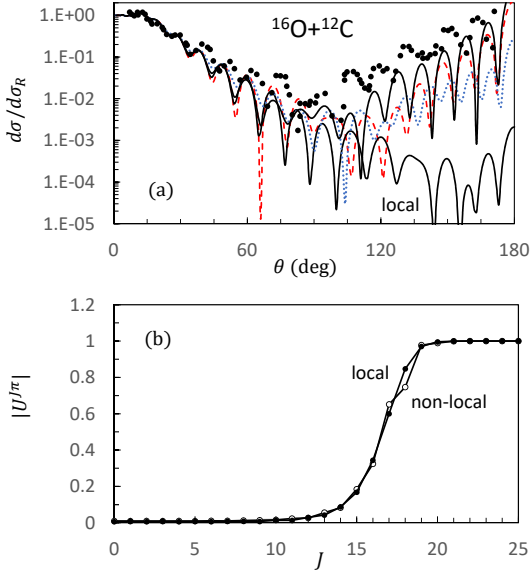,width=7cm}
\caption{$\co$ elastic cross section (divided by the Rutherford cross section) at $\ecm=23.14$ MeV (a).
The data are taken from Ref.\ \cite{VLL89}. The solid lines correspond to the multichannel model.
The dashed lines are obtained with the $^{16}$O ground state only, and the dotted lines with different 
$\ac$ potentials (see text). Panel (b) presents the amplitudes $\vert U^{J\pi}\vert$ of the scattering matrices.}		
\label{fig_c12o16_ela}
\end{center}
\end{figure}

The elastic cross section is presented in Fig.\ \ref{fig_c12o16_ela}(a) in different conditions. The calculation with the local potential provides a fair description of the data up to $\theta \approx 90^{\circ}$, but does not reproduce the
enhancement at large angles. With the non-local term, even if the oscillations at backward angles are not exactly
reproduced, the role of inelastic channels is obvious. The present model, based on the exchange of an $\alpha$
particle during the collision, cannot be expected to be perfect. Other channels are open, such as the neutron or proton
transfer, but are neglected here. There are usually treated by phenomenological potentials involving additional
parameters.

In the $\co$ system, the role of the core-valence potential in more important than in $\ccb$. To assess this
sensitivity, we have also used the $\alpha$ optical potential of Avrigeanu {\sl et al.} \cite{AOR09} (dotted
line in Fig.\ \ref{fig_c12o16_ela}(a)). This is explained by the long range of the $\ac$ potential,
due to the Coulomb term. In this system, using a consistent $\ac$ potential in the entrance and in the
exit channels seems more appropriate.

The amplitude of the scattering matrices $\vert U^{J\pi}\vert$ is displayed in Fig.\ \ref{fig_c12o16_ela}(b).
As for $\ccb$, the calculation with the non-local term provides differences between even and odd partial waves. These differences, however, are weaker than in $\ccb$. As observed in Ref.\ \cite{PMP19}, the even-odd effect
is stronger around the grazing angular momentum ($J\approx 18$).

\section{Discussion of the non-locality}
\label{sec4}
\subsection{Local equivalent potentials}
The effects of the non-locality can be simulated by an equivalent local potential. For the sake of 
simplicity, we assume a single-channel problem. The extension to multichannel systems is simple and does
not modify the conclusions. In a single-channel model, the radial Schr\"odinger equation reads
\begin{align}
&\biggl[-\frac{\hbar^2}{2\mu}\biggl( \frac{d^2}{dr^2}-\frac{L(L+1)}{R^2}\biggr)
+V^{J\pi}(R)-E\biggr]g^{J\pi}(R)\nonumber \\
&\hspace*{1cm} +\int W^{J\pi}(R,R') g^{J\pi}(R') dR'=0,
\label{c1}
\end{align}
and can be replaced by
\begin{align}
&\biggl[-\frac{\hbar^2}{2\mu}\biggl( \frac{d^2}{dr^2}-\frac{L(L+1)}{R^2}\biggr)
+V^{J\pi}_{\rm eq}(R)-E\biggr]g^{J\pi}(R)=0,
\label{c2}
\end{align}
where the equivalent potential $V^{J\pi}_{\rm eq}(R)$ is given by
\begin{align}
&V^{J\pi}_{\rm eq}(R)=V^{J\pi}(R)+\frac{1}{g^{J\pi}(R)}\int W^{J\pi}(R,R') g^{J\pi}(R') dR'.
\label{c3}
\end{align}
This potential depends on the angular momentum, and presents singularities at the nodes of the wave functions.
Thompson {\sl et al.} \cite{TNL89} have proposed to define a smooth, $J$-independent, effective potential by
\begin{align}
&V^{\pi}_{\rm eff}(R)=\frac{\sum_J \omega^{J\pi}(R) V_{\rm eq}^{J\pi}(R)}{\sum_J \omega^{J\pi}(R)},
\label{c4}
\end{align}
where the weight factors $\omega^{J\pi}(R)$ are given by
\begin{align}
&\omega^{J\pi}(R)=(2J+1)(1-\vert U^{J\pi}\vert^2 )\vert g^{J\pi}(R)\vert^2.
\label{c5}
\end{align}
In this way, the influence of the nodes is reduced and the potential (\ref{c5}) does not depend on $J$. However,
the scattering matrices obtained with (\ref{c4}) are not strictly identical to those obtained with (\ref{c1})
or (\ref{c2}). A test with the cross sections must be performed to check the accuracy of the potential (\ref{c4}).

As we expect the equivalent local potentials to depend on parity, the potential (\ref{c4}) is defined for each parity. From
$V^{+}_{\rm eff}(R)$ and $V^{-}_{\rm eff}(R)$, we determine central and parity-dependent potentials as
\begin{align}
&V_0(R)=\frac{1}{2}\bigl(V^{+}_{\rm eff}(R)+V^{-}_{\rm eff}(R)\bigr), \nonumber \\
&V_{\pi}(R)=\frac{1}{2}\bigl(V^{+}_{\rm eff}(R)-V^{-}_{\rm eff}(R)\bigr). 
\label{c6}
\end{align}
The present work, based on rigorous non-local potentials, offers the possibility to investigate the parity potential which,
in general, is phenomenological (see, for example, Ref.\ \cite{LBD95}). Notice that both $V_0(R)$ an $V_{\pi}(R)$
contain real and imaginary components. A parity effect can be also deduced from techniques based on
data inversion \cite{Ma19,PMP19}.

\subsection{Asymptotic form of the non-local kernels}
Here we present some qualitative aspects regarding the non-local kernels, and in particular the overlap kernel
${\cal N}^{J\pi}_{cL,c'L'}(R,R')$. Let us first discuss the overlap functions $N^K_{cc'}(R,R')$ which show up
in Eq.\ (\ref{eq19}). To simplify the presentation, we assume that the valence particle is a neutron in a $s$ state.
In that case, the core-valence wave function tends to
\begin{align}
u_0 (r) \rightarrow C \exp(-k_B r),
\label{d1}
\end{align}
where $k_B$ is the wave number and $C$ the asymptotic normalization coefficient (ANC). 

To develop further, we use the expansion \cite{BG66a}
\begin{align}
&\frac{\exp(-k \vert \pmb{r}_1-\pmb{r}_2 \vert)}{k \vert \pmb{r}_1-\pmb{r}_2 \vert}=\nonumber \\
&\hspace*{1cm}\frac{2}{\pi}
\sum_{\ell}(2\ell+1)i_{\ell}(kr_{<}) k_{\ell}(kr_{>})P_{\ell}(\cos \theta_r),
\label{d3}
\end{align}
where $r_{<}=\min(r_1,r_2)$ and $r_{>}=\max(r_1,r_2)$, and where $\theta_r$ is the angle between $\pmb{r}_1$ and
$\pmb{r}_2$. In this definition, $i_{\ell}(x)$ and $k_{\ell}(x)$ are modified spherical Bessel functions \cite{AS72}.
For large arguments, they tend to
\begin{align}
&i_{\ell}(x)\rightarrow \frac{\exp(x)}{2x}, \nonumber \\
&k_{\ell}(x)\rightarrow \pi \frac{\exp(-x)}{2x}.
\label{d4}
\end{align}
Using the expansion (\ref{d3}) for $u_0 (r)$ and $u_0 (r')$, and using relations (\ref{d2}), we find, for large
$(R,R')$ values
\begin{align}
\frac{u_0 (r)u_0 (r')}{rr'} \rightarrow
\sum_K N^{K,{\rm as }}(R,R')
P_K(\cos \theta_R).
\label{d5}
\end{align}
In the range $\alpha R'<R<R'/\alpha$, the asymptotic kernels are defined by
\begin{align}
&N^{K,{\rm as }}(R,R')= \frac{4}{\pi^2}k_B^2 C^2(-1)^K
\sum_{\ell_1\ell_2}(2\ell_1+1)(2\ell_2+1) \nonumber \\
&\times 
\langle \ell_1 0 \ell_2 0 \vert K 0\rangle^2
k_{\ell_2}(\gamma k_BR)k_{\ell_1}(\gamma k_BR') \nonumber \\
&\times i_{\ell_1}(\alpha\gamma  k_BR)i_{\ell_2}(\alpha\gamma  k_BR'),
\label{d6}
\end{align}
and Eq.\ (\ref{d4}) provides
\begin{align}
N^{K,{\rm as }}(R,R') \sim \frac{1}{R^2R'^2}\exp
\biggl(-\frac{k_B}{1+\alpha}(R+R')\biggr).
\label{d7}
\end{align}
As $\alpha$ is in general close to 1, Eqs.\ (\ref{d6},\ref{d7}) are valid for $ R \approx R'$. Otherwise,
a similar development gives
\begin{align}
	N^{K,{\rm as }}(R,R') \sim \frac{1}{R^2R'^2}\exp
	\biggl(-\frac{k_B}{1-\alpha}\vert R-R'\vert\biggr).
	\label{d7b}
\end{align}
This shows that the non-locality overlap kernel presents an exponential decrease, associated with the
binding energy of the projectile. 

When the angular momentum is not an $s$ wave, the expansion (\ref{d3}) can
be generalized \cite{BG66a}, but this does not change the general trend. In addition, if the transferred particle is
charged, the asymptotic behaviour takes the form
\begin{align}
u_0 (r) \rightarrow C \frac{\exp(-k_B r)}{r^{\eta_B}},
\label{d8}
\end{align}
where $\eta_B$ is the Sommerfeld parameter. The faster decrease can be simulated by using (\ref{d1}) with a
(larger) effective wave number, which simulates Coulomb effects \cite{BG68}. Consequently, Eq.\ (\ref{d6}) remains qualitatively
valid, even for charged transferred particles.

The non-local potentials (\ref{eq14}) also involve kinetic-energy and nucleus-nucleus potential terms. As discussed
in Sec.\ \ref{sec2}.B, the kinetic-energy kernel is directly deduced from a second derivative of the overlap. The asymptotic
behaviour is therefore similar to (\ref{d6}). To determine the potential contribution, we start from definition (\ref{eq19}).
The first term in the potential $V_{Cv}(r')$ is involved in standard DWBA calculations (post form) \cite{Sa83}. For a transferred
neutron, the nuclear contribution in $V_{Cv}$ makes this term short-ranged. The asymptotic behaviour
(\ref{d7}) is therefore modified, with a smaller range. In contrast, the core-core interaction $V_{CC}$ always presents
a Coulomb term. One therefore expects this interaction to be dominant at large distances. 

The non-local potentials in Eq.\ (\ref{eq14}) are obtained from (\ref{eq19}),(\ref{d1}), and angular matrix elements 
\cite{Sa83,SD19}. The procedure is similar to the one followed in DWBA calculations. If the internal angular momenta are
taken as 
$\ell=\ell'=0$, the calculation is simple, as the sum over $K$ contains a single term $K=J$. We have
\begin{align}
{\cal N}^{J\pi}(R,R')=&{\cal J}RR'N^J(R,R')/(2J+1), \nonumber \\
{\cal V}^{J\pi}(R,R')=&{\cal J}RR'\bigl(V^J_{Cv}(R,R')+V^J_{CC}(R,R')\bigr)\nonumber \\
&/(2J+1).
\label{d9}
\end{align}

Equation (\ref{d6}) contains a phase factor $(-1)^J$ at large distances, which shows that the non-local potentials have
opposite signs for even and odd partial waves. This effect arises from the symmetrization of the wave functions
for the core exchange. As shown by von Oertzen \cite{Vo70}, it can be simulated by a local, parity-dependent, potential. An
application to the analysis of the $\ccb$ and of the $\cnb$ systems can be found in Ref.\ \cite{LBD95}.

\subsection{Application to $\ccb$}
\begin{figure}[htb]
	\begin{center}
		\epsfig{file=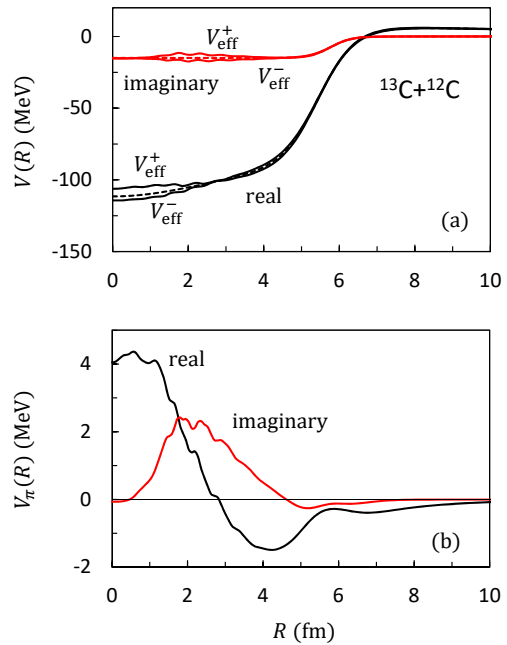,width=6.5cm}
		\caption{Local effective potentials for the $\ccb$ system at $\ecm=7.8$ MeV. In panel (a), the dashed curve
			corresponds to the local potential, and the solid curves to the local equivalent potentials (\ref{c4}).
			Panel (b) displays the parity potential $V_{\pi}(R)$ (\ref{c6}).}
		\label{fig_c12c13_pot}
	\end{center}
\end{figure}

In  Fig.\ \ref{fig_c12c13_pot}, we present the effective potentials for the $\ccb$ system at $\ecm=7.8$ MeV. For the
sake of simplicity, we discuss the single-channel calculation.  Figure \ref{fig_c12c13_pot}(a) shows the local
potential (dashed lines), and the parity-dependent effective potentials (solid lines) obtained with (\ref{c6}). The
elastic cross sections obtained with this effective potential is very close to the original cross section (they
are almost indistinguishable in a figure). As expected, the potentials present small oscillations due to the
nodes in the scattering wave functions. We have repeated the calculations with different numerical conditions 
(channel radius, number of basis functions), and checked that the effective potentials are quite stable.
Figure \ref{fig_c12c13_pot}(b) display the parity potential $V_{\pi}$ [see Eq.\ (\ref{c6})]. The parity effect is important
in the real part as well as in the imaginary part.

\begin{figure}[htb]
	\begin{center}
		\epsfig{file=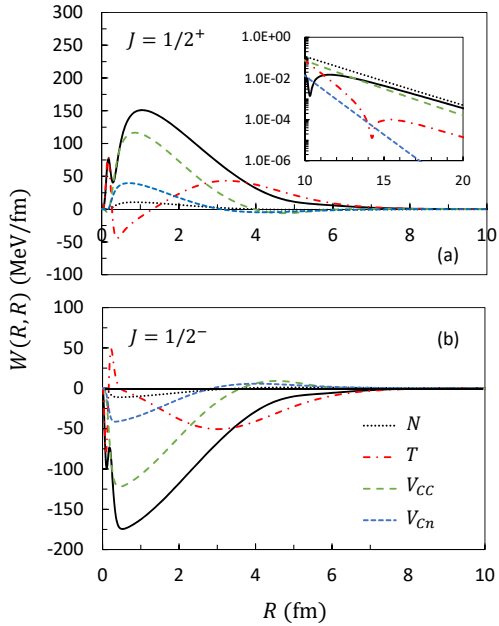,width=6.5cm}
		\caption{Contributions of the overlap, kinetic energy and potentials ($V_{CC}$ and $V_{Cn}$) to the non-local
			kernel (\ref{eq14}) for $R=R'$, and for  $J=1/2^+$ (a) and $J=1/2^-$ (b) in the $\ccb$ system
			($\ecm=7.8$ MeV). The overlap kernel is multiplied by $-\ecm$. For $V_{CC}$, the real part
			is displayed. The inset focuses on the
			long-range part in a logarithmic scale.}
		\label{fig_c12c13_as}
	\end{center}
\end{figure}

In  Fig.\ \ref{fig_c12c13_as}, we display the different contributions in the non-local kernel $W^{J\pi}(R,R')$
(\ref{eq14}) for $R=R'$ at $\ecm=7.8$ MeV. Four terms are present: the overlap, the kinetic energy, and two contributions of the
potential (the core-core and core-valence terms). Partial waves $J=1/2^+\ (L=1)$ and $J=1/2^-\ (L=0)$ are shown in
panels (a) and (b). The change of sign between both parities is confirmed. At short distances, the main contribution
comes from $V_{CC}$, but the kinetic-energy is dominant at large distances. For consistency, the overlap  is multiplied by $-\ecm$, and represents a small contribution. The inset
of panel (a) confirms the asymptotic behaviour. All terms but $V_{Cv}$ have the same exponential behaviour. The
contribution associated with $V_{Cv}$ presents a faster decrease owing to the absence of the Coulomb interaction.

\begin{figure}[htb]
	\begin{center}
		\epsfig{file=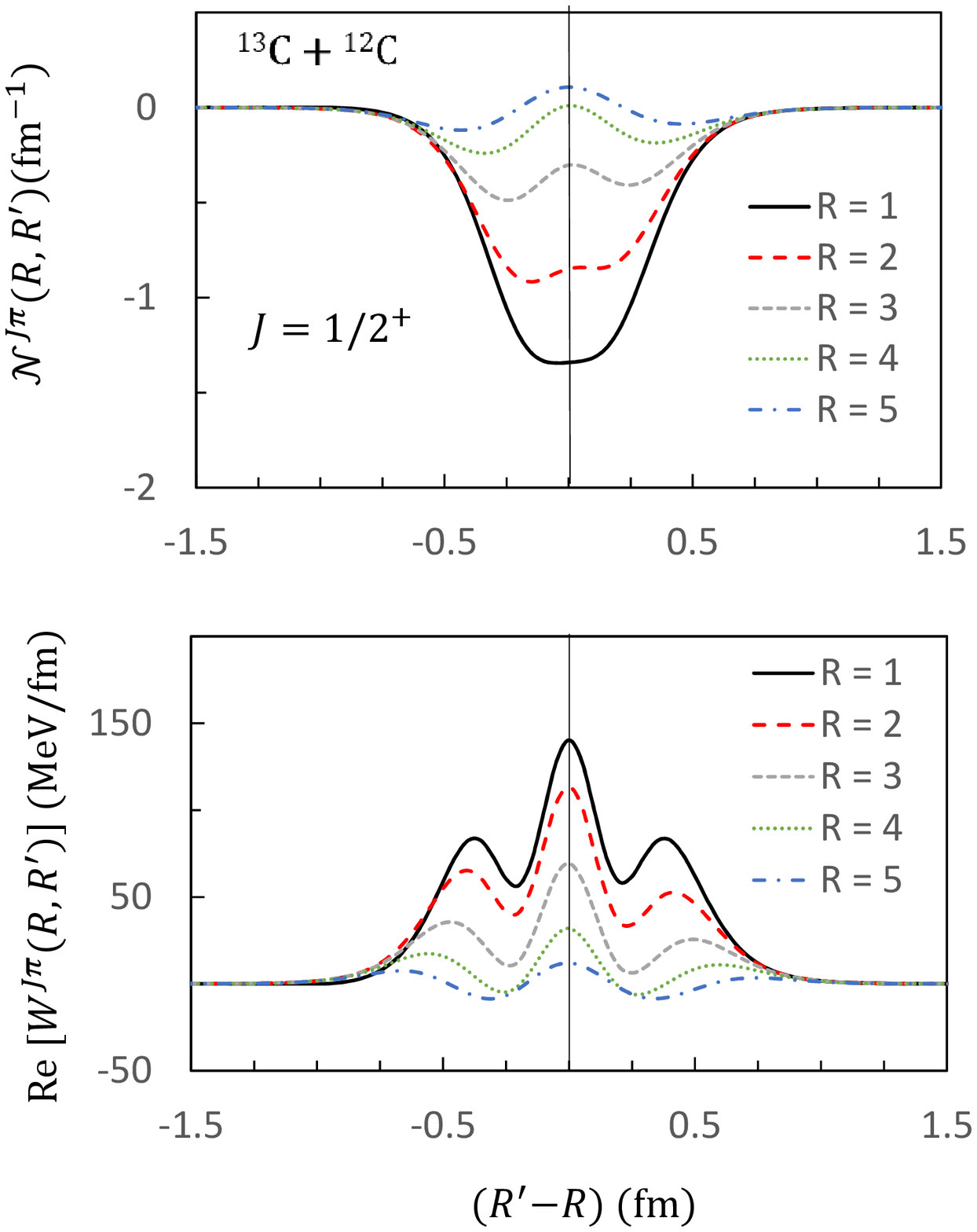,width=6.1cm}
		\caption{Non-local kernels ${\cal N}(R,R')$ (a) and $W(R,R')$ (real part, (b)) for the $\ccb$ system ($J=1/2^+$),
			as a function of $(R'-R)$ [see Eq.\ (\ref{eq14})]. The curves are plotted for different $R$ values.}
		\label{fig_c12c13_nl}
	\end{center}
\end{figure}

In Fig.\ \ref{fig_c12c13_nl}, we analyze the non-locality for $R\neq R'$ and for $J=1/2^+$. We plot the overlap and real-potential
kernels as a function of $(R'-R)$ for various $R$ values. For small $R$ values, the shape of the overlap kernel
is close to a Gaussian, which justifies the Perey-Buck approximation \cite{PB62}. However, for $R>1$ fm, the shape
is more complicated than a Gaussian factor. As expected the main effect of the non-locality comes from $\vert R-R'\vert \lesssim 
1 $ fm.

\subsection{Application to $\co$}
\begin{figure}[htb]
	\begin{center}
		\epsfig{file=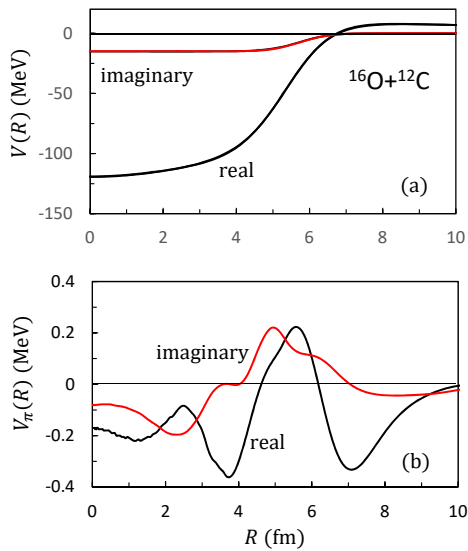,width=6.1cm}
		\caption{Local effective potentials for the $\co$ system at $\ecm=23.14$ MeV. In panel (a), the 
	local potential and the equivalent potential (\ref{c4}) are superimposed.
	Panel (b) displays the parity potential $V_{\pi}(R)$ (\ref{c6}).}
		\label{fig_c12o16_pot}
	\end{center}
\end{figure}

The local effective potentials (\ref{c4}) for $\co$ at $\ecm=23.14$ MeV are presented in Fig.\ \ref{fig_c12o16_pot}(a). Again, we limit
the discussion to the single-channel system. Since the depth
of these potentials is rather large, the differences with the exact local potential (\ref{eq13}) are small,
indistinguishable at the scale of the figure.
Figure \ref{fig_c12o16_pot}(b) shows the parity potential (\ref{c6}). As for $\ccb$, these potentials
present oscillations due to the nodes of the scattering wave functions. Although the amplitude is small, the effect
of the parity potential extends to large distances. This explains that this potential provides a large backward-angle
enhancement of the cross section (see Fig.\ \ref{fig_c12o16_ela}). The cross sections provided by the effective potential
(\ref{c4}) are very close to the original cross sections of  Fig.\ \ref{fig_c12o16_ela}.

\begin{figure}[htb]
	\begin{center}
		\epsfig{file=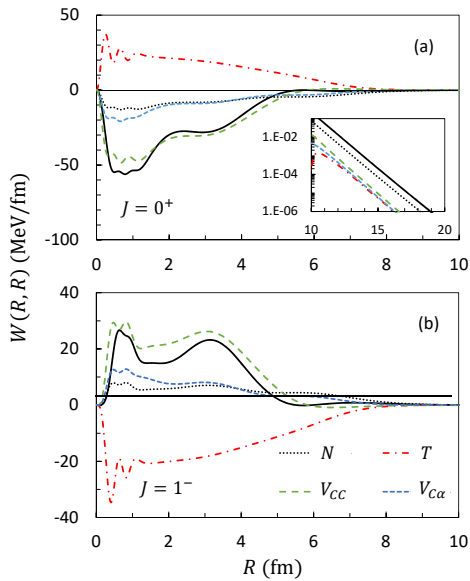,width=6.1cm}
		\caption{Contributions of the overlap, kinetic energy and potential ($V_{CC}$ and $V_{C\alpha}$) to the non-local
			kernel (\ref{eq14}) for $R=R'$, and for  $J=0^+$ (a) and $J=1^-$ (b) in the $\co$ system. The real part of $V_{CC}$ is shown. The inset focuses on the
			long-range part in a logarithmic scale.}
		\label{fig_c12o16_as}
	\end{center}
\end{figure}

The decomposition of the non-local kernel (\ref{eq14}) in different terms is presented in Fig.\ \ref{fig_c12o16_as} for
$J=0^+$ (a) and $J=1^-$ (b). The dominant contribution comes from the core-core potential $V_{CC}$. There is a cancellation
effect of the overlap, kinetic-energy and $V_{C\alpha}$ contributions. At large distances the decrease of the potentials
(inset of Fig.\ \ref{fig_c12o16_as}(a) is much faster than in $\ccb$ (see Fig.\ \ref{fig_c12c13_as}). This is explained
by Eq.\ (\ref{d7}) since $k_B$ is much larger in $^{16}$O than in $^{13}$C (and coefficient $\alpha$ is smaller).

\begin{figure}[htb]
	\begin{center}
		\epsfig{file=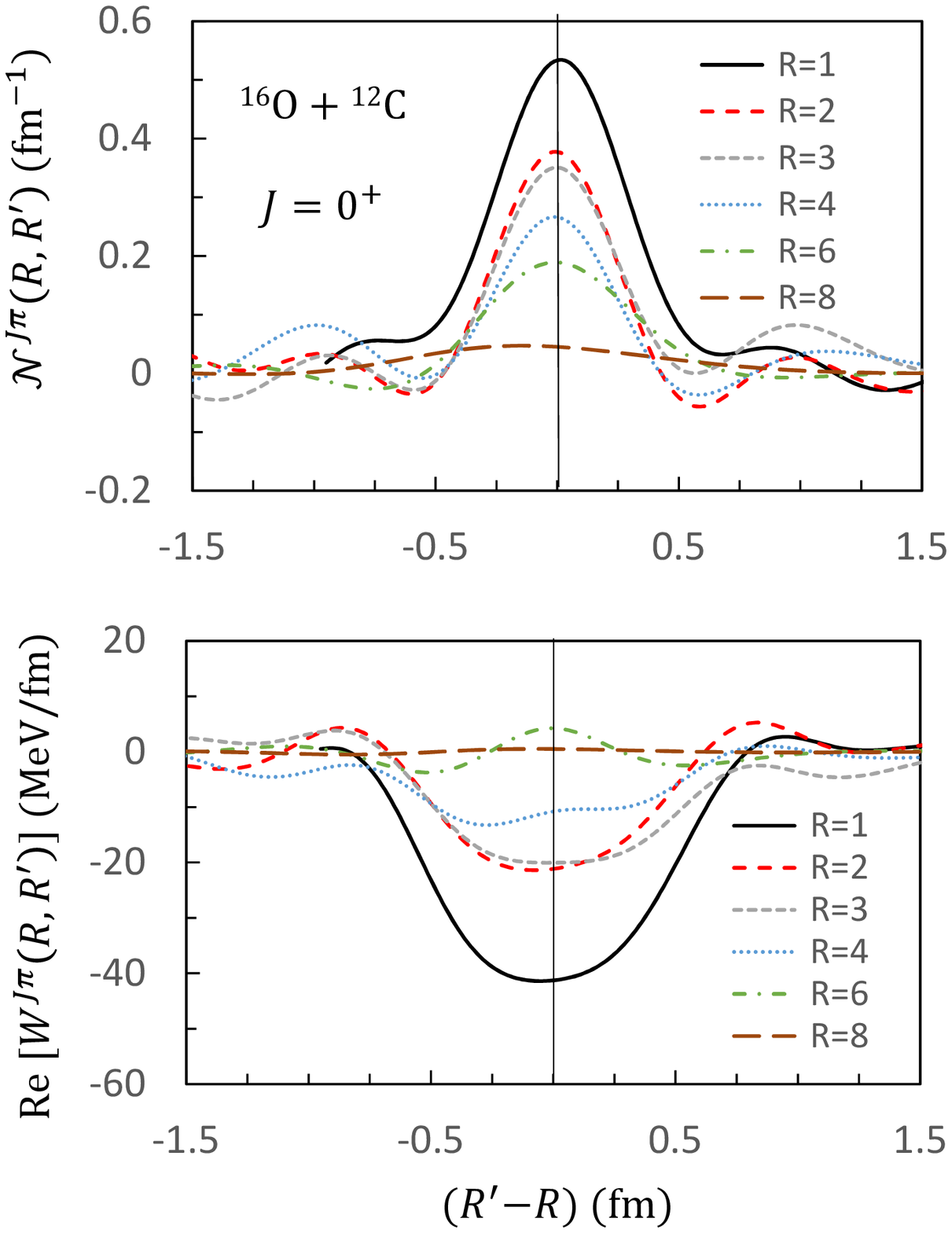,width=6.1cm}
		\caption{Non-local kernels ${\cal N}(R,R')$ (a) and $W(R,R')$ (real part, (b)) for the $\co$ system ($J=0^+$),
	as a function of $(R'-R)$ [see Eq.\ (\ref{eq14})]. The curves are plotted for different $R$ values.}
		\label{fig_c12o16_nl}
	\end{center}
\end{figure}

The non-locality for $R\neq R'$ is illustrated in Fig.\ \ref{fig_c12o16_nl} for $J=0^+$. As for $\ccb$, the shape of 
the overlap kernel is close to a Gaussian for small $R$ values, but deviates when $R$ increases.

\section{Conclusion}
\label{sec5}
Our main goal is an exploratory study of a rigorous method to treat exchange effects in nucleus-nucleus scattering.
Starting from a three-body model, we have derived local and non-local kernels in a coupled-channel
formalism. This represents a natural extension of the three-body CDCC method. The coupled-channel system,
involving non-local potentials, is solved with the $R$-matrix theory, associated with the Lagrange-mesh
technique. This permits fast and accurate calculations of the scattering matrices and of the cross sections.

We have applied the formalism to typical reactions: $\ccb$, $\cnb$ and $\co$, which illustrate the transfer of a neutron,
of a proton, and of an $\alpha$ particle, respectively. In each case we started from a core-core optical potential,
taken from the literature, and
which fits $\cca$ elastic-scattering data. As expected, exchange effects lead to a backward-angle enhancement of the
elastic cross sections. For the core-valence potential, we have two ``natural" choices: either it is fitted on the
spectroscopic properties of the heavy particle, or it is fitted on scattering properties. The first choice
guarantees the symmetry of the scattering matrix, since the same Hamiltonian is used in the entrance and exit channels.
With our examples, we have however shown that both options provide similar cross sections. This can be explained
by the weak contribution of this potential to the non-local kernel.

We have shown in Sec.\ \ref{sec4} that the bound-state wave functions of the $C+v$ system need to be accurately determined
up to large distances. If not, the calculation of the various kernels is inaccurate, and the scattering matrices
are unstable. For this reason, CDCC calculations involving continuum states raise numerical difficulties with
pseudostates as they tend to zero at very large distances only. The use of bins would be preferable. In our
applications, however, the binding energies of the projectile are large, and no strong continuum effects are
expected.

The present work opens the path to more ambitious calculations, such as the $\abe$ system, where
exchange effects could affect the existing calculations \cite{OKK09}. Also it could be extended to more complicated
systems, such as $\dbe$ or $\cnc$ which require a four-body theory. Another application of the formalism
deals with Coupled Reaction Channel (CRC) calculations where similar non-local kernels are involved \cite{Sa83}.

\section*{Acknowledgments}
This work was supported by the Fonds de la Recherche Scientifique - FNRS under Grant Numbers 4.45.10.08 and J.0049.19.
It benefited from computational resources made available on the Tier-1 supercomputer of the 
F\'ed\'eration Wallonie-Bruxelles, infrastructure funded by the Walloon Region under the grant agreement No. 1117545. 

\onecolumngrid
\appendix
\section{Calculation of the non-local kinetic energy}
\label{appendixa}
The kinetic-energy kernel ${\cal T}^{J\pi}_{cL,c'L'}$ is implicitly defined by 
\begin{align}
	\int {\cal T}^{J\pi}_{cL,c'L'}(R,R')g^{J\pi}_{c'L'}(R')dR'=&
	R\langle \varphi^{JM\pi}_{cL} \vert T_{\pmb{R}} \vert
	\varphi^{JM\pi}_{c'L'}\dfrac{g^{J\pi}_{c'L'}(R')}{R'}\rangle,
	\label{eqA1}
\end{align}
where the integration is performed over $\pmb{r}$ and $\Omega_R$. An explicit expression for this kernel 
can be obtained by, first, writing the kinetic-energy operator $T_{\pmb{R}}$ in spherical coordinates, 
\begin{align}
	T_{\pmb{R}}=-\dfrac{\hbar^2}{2\mu R}\dfrac{d^2}{dR^2}R+\dfrac{\hat{L}^2_{\pmb{R}}}{2\mu R^2},
\end{align}
where $\hat{L}_{\pmb{R}}$ is the orbital angular momentum associated with the coordinate $\pmb{R}$. The radial part of $T_{\pmb{R}}$ can be moved outside the matrix element since the function $\varphi^{JM\pi}_{cL}$ is independant of $R$ and since the matrix element does not involve any integration over the coordinate $R$. Besides, since the operator $\hat{L}^2_{\pmb{R}}$ is Hermitian, it can be applied, with a trivial effect, on the bra.
We therefore obtain 
\begin{eqnarray}
	R\langle \varphi^{JM\pi}_{cL} \vert T_{\pmb{R}} \vert
	\varphi^{JM\pi}_{c'L'}\dfrac{g^{J\pi}_{c'L'}(R')}{R'}\rangle&=&-\dfrac{\hbar^2}{2\mu}\left[\dfrac{d^2}{dR^2}-\dfrac{ L(L+1)}{R^2}\right] R\langle \varphi^{JM\pi}_{cL} \vert 
	\varphi^{JM\pi}_{c'L'}\dfrac{g^{J\pi}_{c'L'}(R')}{R'}\rangle\\
	&=& \int \left\{-\dfrac{\hbar^2}{2\mu}\left[\dfrac{\partial^2}{\partial R^2}-\dfrac{L(L+1)}{R^2}\right]{\cal N}^{J\pi}_{cL,c'L'}(R,R')\right\}g^{J\pi}_{c'L'}(R')dR',
	\label{eqA6}
\end{eqnarray}
where the definition~\eqref{eqadd} of the norm kernel and the Leibniz integral rule have been used. By comparison of \eqref{eqA1} and \eqref{eqA6}, we get the relation~\eqref{eq20}, linking the kinetic-energy and norm kernels,
\begin{align}
	{\cal T}^{J\pi}_{cL,c'L'}(R,R')=-\frac{\hbar^2}{2\mu}
	\biggl[\frac{\partial^2}{\partial R^2}-\frac{L(L+1)}{R^2}\biggr]
	{\cal N}^{J\pi}_{cL,c'L'}(R,R').
\end{align}
\twocolumngrid
%

\end{document}